\documentclass[11pt]{article}
\usepackage{graphicx}
\usepackage{amssymb}

\begin{document}

\title{Extragalactic circuits, transmission lines, and CR particle acceleration}

\author{Philipp P. Kronberg (1) \& Richard V.E. Lovelace (2)\\
(1) University of Toronto, Department of Physics, 60 St George Street,\\ Toronto M5S 1A7 Canada: email; philkronberg@gmail.com\\
(2)  Cornell University, Department of Astronomy, Ithaca NY 14853 USA}

\date{}

\maketitle

\begin{abstract}
A non-negligible fraction of
a Supermassive Black Hole's (SMBH) rest mass energy gets transported into extragalactic space by a remarkable process in jets which are incompletely understood.  What are the physical processes which transport this energy? 
   It is likely that the  energy flows electromagnetically, rather than via a particle beam flux.  The deduced electromagnetic fields may produce particles of energy  as high as  $\sim 10^{20}$ eV. 
   The energetics of SMBH accretion disk models and the electromagnetic energy transfer imply that a SMBH should generate a $10^{18} - 10^{19}$  Amp\`eres current close to the  black hole and its accretion disk.
   We describe the so far best  observation-based estimate of the magnitude of the current flow along the axis of  the  jet extending from the nucleus of the active galaxy in  3C303.   
    The current is measured to be $I \sim 10^{18}$ Amp\`eres at $\sim 40$ kpc away from the AGN.  
       This indicates that organised  current
flow  remains intact  over multi-kpc distances.
    The electric current $I$ transports electromagnetic  power into free space, $P = I^{2}Z$, where $Z \sim 30$ Ohms is related to the impedance of free space, and this points to the existence of  cosmic electric circuit.
    The associated electric potential drop, $V=IZ$, is of the order
of that required to generate Ultra High Energy Cosmic Rays (UHECR).
     We also explore  further implications, including disruption/deflection of the power flow and also why such measurements, exemplified by those on 3C303, are currently very difficult to make and to unambiguously interpret. This naturally leads to the topic of how such measurements can be extended and improved in the future. 
We describe the analogy of electromagnetically dominated jets with transmission lines.  High powered jets {\it in vacuo} can be understood by approximate analogy with  a waveguide. The importance of inductance, impedance, and other laboratory electrical concepts are discussed in this context.   To appear in Proc. 18th International Symposium on Very High Energy Cosmic Ray Interactions (ISVHECR2014), CERN, Switzerland

\end{abstract}

\section{Introduction}

\label{sec:1}

\label{intro} Astrophysical jets of collimated energy flow have many forms and different progenitors $-$ ranging from the jets of
accreting protostars to the relativistic jets from accreting SMBHs.
    The physical basis of the energy flow in the relativistic jets
remains uncertain.
    Here we focus on the most massive end of this scale, the
 jets from SMBHs.
    The jet power is assumed to be predominantly electromagnetic as proposed in  \cite{b1} coming from the power released from the accretion disk  and/or from the spin-down of a rotating black hole  \cite{b5}. 
      General relativistic simulations of such systems can be found in \cite{b6},  \cite{b7}, and  \cite{b8}.
We focus on two aspects of these ``ultimate energy transfer'' machines, which produce the largest post-inflation energy transfers in the universe. 

First we demonstrate an  ``experiment on the sky'' that shows, quantitatively, how SMBH's can transfer enormous amount of
electromagnetic energy from its vicinity into the intergalactic medium\cite{b2}. 
     The second is to propose, based on models and recent astrophysical measurements,
how these remarkably collimated jets can be considered as giant electro-magnetic circuits $-$ on multi-kpc scales in intergalactic space \cite{b3},~\cite{b4}. 
        The energetics and the structure of these jets also make them capable, in principle, of accelerating cosmic rays up to $\sim 10^{20}$ eV. If so, this would make SMBHs and their accretion disk/jet systems favoured acceleration sites for producing UHECRs.

\section{Calorimeters of the energy released from supermassive black holes }

\label{sec:2}

We compare two samples of extended radio galaxy lobes were that were imaged 
and analysed \cite{b2}. They are (i) Giant Radio Galaxies (GRG) {\it outside} of galaxy clusters, where the ambient gas density is $\lesssim 10^{-5} {\rm cm}^{-3}$. 
     Here, the acceptance criterion was a minimum projected size of $0.6$ Mpc. This includes some of the largest radio galaxies known, having projected sizes of up to a few mega-parsecs, i.e. of order a galaxy - galaxy spacing in the large scale structure (LSS) of the Universe. The second sample (ii) consists of extended radio sources located {\it within} $\approx 150$ kiloparsecs of a cluster centre where the ambient gas density is $\gtrsim 10^{-3} {\rm cm}^{-3}$.   

A plot of the energy content of the radio emitting lobes vs. the largest angular size (LAS) of the system reveals a remarkable result. In Fig. 1 we see that the upper envelope of energy content of the GRG radio lobes in sample (i) is within ${\sim 2}$ orders of magnitude of the (gravitational) formation energy of a $10^8 {\rm M}_\odot $ black hole. This means that the energy content of the CR's and magnetic fields in these large magnetised radio lobes (fed by the jet over $10^7 - 10^8$ yr) is a non-negligible fraction of the original gravitational energy ``reservoir''.  Stated differently, the lobes represent captured energy, fed by the jets and which has not been ``lost'' as $PdV$ work against a significant ambient confining pressure.   

Sources in the second sample (ii) have had to expand against the  counter-pressure of the intracluster gas, which is independently measurable from X-ray observations. Their upper envelope of total energy content turns out to be lower than that for the GRG's by just the amount of the $PdV$ work against the intra-cluster gas. The upper envelope of the latter, cluster-internal, group, if raised by this $PdV$ work, would approximately match that of the GRG sample (i).

The above discussion sets the stage for the subsequent steps: From the above we can conclude the age, total energy flow rate, and the (very low) ambient non-relativistic gas density around the jets.

\begin{figure*}
\centering
\includegraphics[width=12cm]{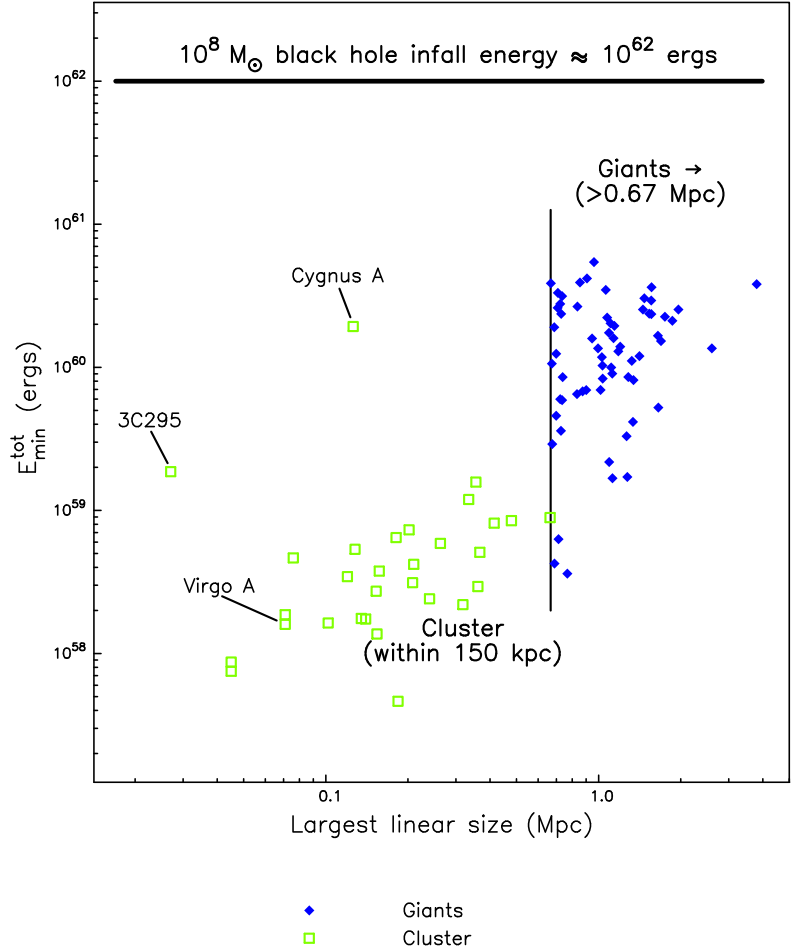}
\caption{Similar to panel 2 of Fig 1 in \cite{b2}. Estimates of the total radio lobe energy content for samples (i) and (ii) of extended radio sources described in the text. Estimated total energy content is plotted against largest linear size (LLS) in Mpc. The formation energy of a $ 10^8 M_\odot $ SMBH is shown for comparison. Blue diamonds represent Giant radio galaxies having projected LLS $> 0.67 $ Mpc, and green squares represent extended radio galaxies within 150 kpc (projected) of the centre of a cluster}
\label{Fig1}
\end{figure*}

\section{Estimate of the jet current}

\label{sec:3}

At the higher resolution of $0.3^{\prime\prime}$ the most extended of the three
jet knots are resolved at 3 frequencies in total intensity and linear 
polarization. This was sufficient to permit measurement  of the Faraday rotation, and Faraday rotation gradients \cite{b3}. The result was $~\sim 0~ {\rm radians /m}^2$ per kpc gradient along the jet axis,
compared with a transverse Rotation Measure (RM)  gradient of $\sim + 10~ {\rm radians /m}^2$ per kpc. 
      If this transverse RM gradient is the signature 
of an axial current in the jet then it should change sign on either side of the jet axis. In the observed image any Faraday rotation in the direction of 3C303 contains 
a foreground component from the Milky Way plus any intergalactic magneto-plasma around the 3C303 system, {\it and}
between it and our Milky Way. To correct for this, we subtracted the average RM of a number of nearby (in sky position)
extragalactic sources. Fortunately this correction is small at 3C303's location on the sky. After applying it, the RM variation
across the jet crosses zero on the jet's axis, within the errors.  This is the signal which is consistent with an axial current.  The implied 
value of the current is  $ 3 \times 10^{18}$ A. The sign of ${\bf \nabla}( RM)$ corresponds to a positive current directed, in this case,
{\it away} from the galaxy nucleus. Further details can be found in \cite{b3}.

\section{The puzzle of the apparent jet disruption and expansion into a radio lobe}

\label{sec:4}

At the apparent disruption point at the end of the 3C303 jet, a lobe proceeds further, but  
in a transverse direction. At $\lesssim 1^{\prime\prime}$ resolution, it appears to 
contain sub-blobs whose low Faraday rotation fluctuation is comparable to that in the jet. This indicates a very
low internal thermal plasma content, as is also the case for the cocoon around the jet. Further, these sub-blobs also appear to contain 
highly ordered magnetic fields which are coherent on the scale of the blobs. The ``disrupter'' in this case is tentatively identified 
as a faint, blue optically visible ``cloud'' that was observed near the limit of detectability (``object G'')  \cite{b9}.
The magneto-plasma effects of such a major jet disruption warrant a separate analysis, and are not discussed further here.

\section{Jets as electric circuits}

\label{sec:5}

If the observational evidence of a net axial current in the jet
of 3C303 is borne out,  and  comparable currents are observed
in other systems, the implications are profound. 
    The existence of an axial current on these dimensions suggests that extragalactic currents flow on  the scale of the lobes. 
      The initial diameter of the jet
is a few times the radius of the SMBH accretion disk radius (\cite{b10}) and is thus difficult to resolve transversely. 
     The power flow in the jet is carried by electromagnetic fields supported
by a very low density plasma.   For such a jet, the natural impedance
 is determined by 
the permeability, $\mu_0 $, and permittivity, $\epsilon_0 $ of free space.
Specifically, the jet impedance is related to the impedance of free space,
$Z\approx (4\pi)^{-1}\sqrt{\mu_0/\epsilon_0} \approx 30~\Omega$  \cite{b4}.   The fundamental quantities $\mu_0$ and $\epsilon_0$ also determine
the order of magnitude of propagation speed of disturbances in the jet, namely, 
$\sim c=(\mu_0\epsilon_0)^{-1/2}$.

Where there is a plasma containing freely moving charges, however rarified, $Z$ can be complex, that is, can have a reactive component. 
Finally, a Poynting flux contains an e.m.f., which is normal to the propagation direction. 
For a SMBH  driven jet, the voltage $V$  at the point of launching
is of the order of $10^{20}$ V \cite{b1}. Thus $V = IZ$, and the 
corresponding power flow rate is $P = I^2 Z$,  as in a laboratory electric circuit. 

The jet will necessarily have a return current which is diffuse and located
away from the jet axis.
The inferred outgoing and return currents will repel each other, and the return current must in this case flow outside the zone defined by the instrumental resolution. Otherwise, given our image linear resolution of $\sim 0.5 $kpc, 
it would cancel a unidirectional current measured within the $\sim 1^{\prime\prime}$ beam (\cite{b11}).

\section{Extension to the analogy of transmission lines}

\label{sec:6}

In the late 1800's the propagation of a telegraph signal over a long transmission line was a new achievement, and an important application of electromagnetic theory. In some ways is was a forerunner 
for our understanding of Poynting flux jets. A pioneer of the electrical analysis of telegraphy was Oliver Heaviside (1893, \cite{b12}). His ``telegrapher's equations'' have application to some phenomena of Poynting flux jets described here \cite{b4}. The telegrapher's equations can be combined to give wave equations which describe 
the time and space perturbations in a transmission line of voltage $\Delta V $ and current $\Delta I$ in a Poynting Flux jet \cite{b4}. They also determine the phase velocity $v_\varphi $ in the line, which in turn depends on the wave impedance of the jet $Z= \sqrt {L /C} $  where $L $, $C$ are its inductive and capacitive components.  

A  well-known  laboratory example is a transmission line with
an intrinsic impedance $Z_0$ connected to a possibly complex load impedance $Z_L$. 
   The standing wave pattern along the $z $-axis due to the oppositely propagating incident and reflected waves from the load can be used to calculate the (mismatched), complex impedance of the load, $Z_L$. 
   This may be one explanation for the regular emission maxima and minima seen along many collimated SMBH-powered jets. Lovelace \& Ruchti \cite{b13} show a calculation of the transmission line voltage, $V_0$, which is the effective potential drop across the line, for an axial current flow  $V_0 / Z_0$. Here, $Z_0 $ is the impedance of the Poynting flux jet (\cite{b4}). 

The resulting sum of outgoing and reflected voltage waves causes the ratio $B / E $ to be perturbed, and it changes quasi-periodically along the $z-$axis of propagation. This has some similarity to the situation of two counter-propagating waves in a laboratory transmission line experiment which can be used to measure an unknown impedance, $Z_L$. The main difference is that the light travel time over a path segment $\delta z $ having low $B/E $ ratio can be long, so that a proton can dwell long enough in a unidirectional $E-$field to reach very high energies via $E_{\Vert}$ acceleration.

\section{Conditions for ``magnetic insulation'' and\\
VHECR/UHECR acceleration in these\\ ``ultimate'' jets}

\label{sec:7}

For our purposes here the ratio $|B| / |E| $ is important: It can represent the degree of ``magnetic insulation'' along the $z$ - axis  \cite{b14}. In the above example (Fig. 2) the path segments ($\Delta z $) in which
$|E| / |B| \gtrsim 1$ ({\it reduced} magnetic insulation) are traversed in a time $\approx \Delta z  $ / $\Delta v_{\varphi }$ ($\Delta z $ not to be confused with an interval of cosmological redshift).

\begin{figure*}
\centering
\includegraphics[width=12cm]{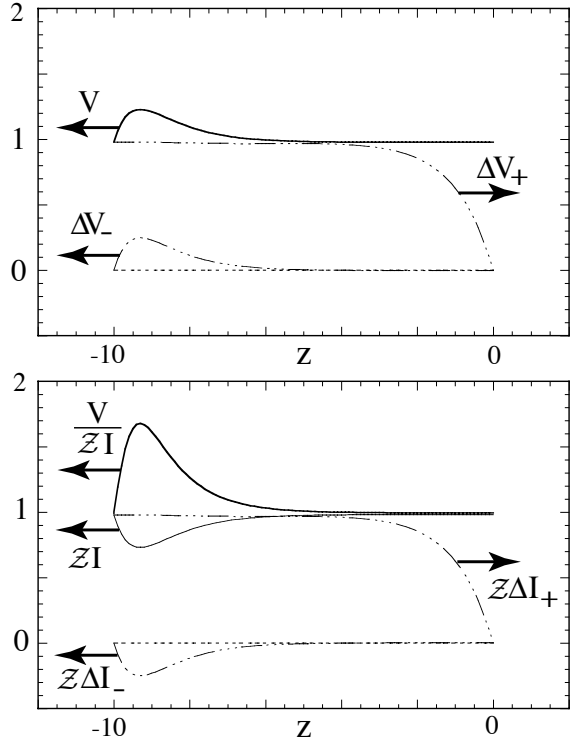}
\caption{Reflection of an incident wave 
($\Delta V_+,~\Delta I_+$)
off of a stationary load of impedance ${\cal Z}_L$
and the reflected wave ($\Delta V_-,~\Delta I_-$) for a
transmission line with an inductive  load at $z=0$.   
 The net voltage is
 $V = \Delta V_+ +\Delta V_-$ and the net current  
 $I= \Delta I_+ +\Delta I_-$.  The top panel shows 
 the voltages and the bottom panel the currents as
 well as the  ratio $V({\cal Z}I)^{-1}$.  
   If this ratio is larger than a critical value, ${\cal C}$,
then the magnetic insulations will breakdown.
 For this example, ${\cal Z}_L={\cal Z} -i\omega {\cal L}$, where
 ${\cal L}$ is the inductance of the load. }
\end{figure*}

During this interval both positive and negatively charged particles can ``dwell'' in a zone where $|E| \gtrsim |B|$, the criterion for CR acceleration along the $z $-axis. In an example of an extragalactic scale jet, $\Delta z = 1 $ kpc corresponds to a dwell time interval $\delta t \sim 3000$ yr for CR acceleration to occur.

\section{What is needed for more jet current measurements, and how the instrumental  capabilities can be extended}

\label{sec:8}

Radio observations are important here, and they require the greatest possible angular resolution and sensitivity, as well as the ability to produce Faraday RM images $--$ which requires observations at two or more 
suitably chosen frequencies. Generally speaking there are two different regimes of angular resolution. The first applies to (i) $\gtrsim $kpc-scale, extended extraglactic jets. This is the subject of my presentation today.

A second and third class (ii) and (iii), not elaborated upon here, are images of the early ``launching'' phase of a jet, close to the SMBH. This happens on scales of parsecs and less, where the resolution needed is $\lesssim $ 
milliarcseconds. The magnetic field structure within the jet-related plasma is a challenge here too, perhaps even more so, than 
for systems similar to the 3C303 jet. The radio instrumental approach for such measurements is Very Long Baseline Interferometry (VLBI) with radio telescope separations ranging from hundreds to thousands of km. This includes maximum separations 
of $\gtrsim 20,000$km in space-VLBI. A pioneering instrument of the latter kind was the Japanese VSOP project having one satellite with an orbital apogee of 22,000 km. Future generations are being developed and planned, 
such as the Russian RADIOASTRON multi-satellite VLBI. In the same context it is appropriate to mention a third (iii), future class of multi-element imaging {\it optical} interferometers: These are Earth-based
interferometers that could achieve comparable resolutions to radio space-VLBI,  $-$ though not with Faraday rotation which is $\propto \lambda^2 $. 

The example of 3C303 (class (i)) suggests that on kpc scales source-intrinsic RM variations due to current patterns in the jet are small, probably $\lesssim 20 \deg $. For small RM variations this requires that such experiments be carried out at frequencies $\lesssim 1 $ GHz for extended jet-lobe radio sources. In this frequency range the needed sub-arcsecond resolution dictates interferometer array dimensions up to a few hundred km. Up to now, 
there are few instruments that that can simultaneously satisfy the criteria of  resolution, sensitivity and lower frequency ($\ll 1 $ GHz capability. Candidates would be a further developed European LOFAR array, 
or perhaps a further development of the existing LWA in the US. An enhanced VLA that was previously proposed with a few hundred km dimension would have been a powerful candidate, but this proposed enhancement was not funded.

\section{Conclusion}

The results and methods briefly described here illustrate an important ongoing development in the merging of laboratory high energy particle physics, with the physics 
of magnetic fields and cosmic rays in space. In the latter, VHECR and UHECR instruments are detecting CR nuclei up to $\sim 10^{20} $ eV, along with VHE $\gamma $ rays, neutrinos
and other sub-atomic species. The overlap between laboratory- and space-based experiments and measurements are destined to grow even more as enhancements and successors to the LHC
appear, and with the future expansion of several high energy detectors of sub-atomic particles and $\gamma $ rays from space.

\section*{Acknowledgments}

We are grateful to the late Stirling Colgate for many illuminating discussions 
on SMBHs and cosmic ray physics in several collaborations. 
PPK acknowledges support of a Discovery Grant from the Natural Sciences and Engineering Research 
Council of Canada (NSERC), the University of Toronto, and the 
Alexander von Humboldt Stiftung, (Germany) for financial support of related research over several years. 
I am also grateful to the U.S. Department of Energy and Los Alamos National Laboratory (LANL) for support while a long term, 
part time Visiting Scholar at LANL.


\end{document}